\documentclass[aps,twocolumn,pra,longbibliography,showpacs,amsmath,amssymb,superscriptaddress,reprint]{revtex4-1}
\usepackage[utf8]{inputenc}

\usepackage[colorlinks=true,citecolor=blue]{hyperref}
\usepackage{soul}
\usepackage{amsfonts}
\usepackage[english]{babel}
\usepackage{graphicx}
\usepackage{dcolumn}
\usepackage{bm}
\usepackage{siunitx}

\newcommand{\kBT}{k_\text{B}T}

\begin{document}

\title{Phase-tunable thermal rectification in the topological SQUIPT}

\author{Lennart Bours}
\affiliation{NEST, Istituto Nanoscienze--CNR and Scuola Normale Superiore, Piazza San Silvestro 12, 56127 Pisa, Italy}
\author{Bj\"orn Sothmann}
\affiliation{Theoretische Physik, Universit\"at Duisburg-Essen and CENIDE, D-47048 Duisburg, Germany}
\author{Matteo Carrega}
\affiliation{NEST, Istituto Nanoscienze--CNR and Scuola Normale Superiore, Piazza San Silvestro 12, 56127 Pisa, Italy}
\author{Elia Strambini}
\affiliation{NEST, Istituto Nanoscienze--CNR and Scuola Normale Superiore, Piazza San Silvestro 12, 56127 Pisa, Italy}
\author{Alessandro Braggio}
\affiliation{NEST, Istituto Nanoscienze--CNR and Scuola Normale Superiore, Piazza San Silvestro 12, 56127 Pisa, Italy}
\author{Ewelina M. Hankiewicz}
\affiliation{Institute for Theoretical Physics and Astrophysics, University of W\"urzburg, Am Hubland, D-97074 W\"urzburg, Germany}
\author{Laurens W. Molenkamp}
\affiliation{Experimentelle Physik 3, Physikalisches Institut, University of W\"urzburg, Am Hubland, D-97074 W\"urzburg, Germany}
\author{Francesco Giazotto}
\affiliation{NEST, Istituto Nanoscienze--CNR and Scuola Normale Superiore, Piazza San Silvestro 12, 56127 Pisa, Italy}

\date{\today}

\begin{abstract}
We theoretically explore the behavior of thermal transport in the topological SQUIPT, in the linear and nonlinear regime. The device consists of a topological Josephson junction based on a two-dimensional topological insulator in contact with two superconducting leads, and a probe tunnel coupled to the topological edge states of the junction. We compare the performance of a normal metal and a graphene probe, showing that the topological SQUIPT behaves as a passive thermal rectifier and that it can reach a rectification coefficient of up to $~ 145\%$ with the normal metal probe. Moreover, the interplay between the superconducting leads and the helical edge states leads to a unique behaviour due to a Doppler shift like effect, that allows one to influence quasi-particle transport through the edge channels via the magnetic flux that penetrates the junction. Exploiting this effect, we can greatly enhance the rectification coefficient for temperatures below the critical temperature $T_\text{C}$ in an active rectification scheme.
\end{abstract}

\maketitle

\section{Introduction}
As electronic circuits shrink down to the nanoscale, the management of heat becomes evermore important\cite{Keyes2005,Mannhart2010,Sanchez2014}. This is especially true for quantum technologies, where fragile quantum states can easily be corrupted through unwanted interactions with a `hot' environment\cite{Spilla2014,Spilla2015}. Simultaneously, control over heat flows at the nanoscale prompts the attractive prospect of on-chip cooling~\cite{Edwards1995,Pekola2000,Miller2006,Giazotto2006,Chowdhury2009,Vischi2018} which could make the application of low temperature quantum technologies considerably cheaper, or could be used further improve operating efficiencies.

\begin{figure}[hbtp]
\centering
	\includegraphics[width=0.8\columnwidth]{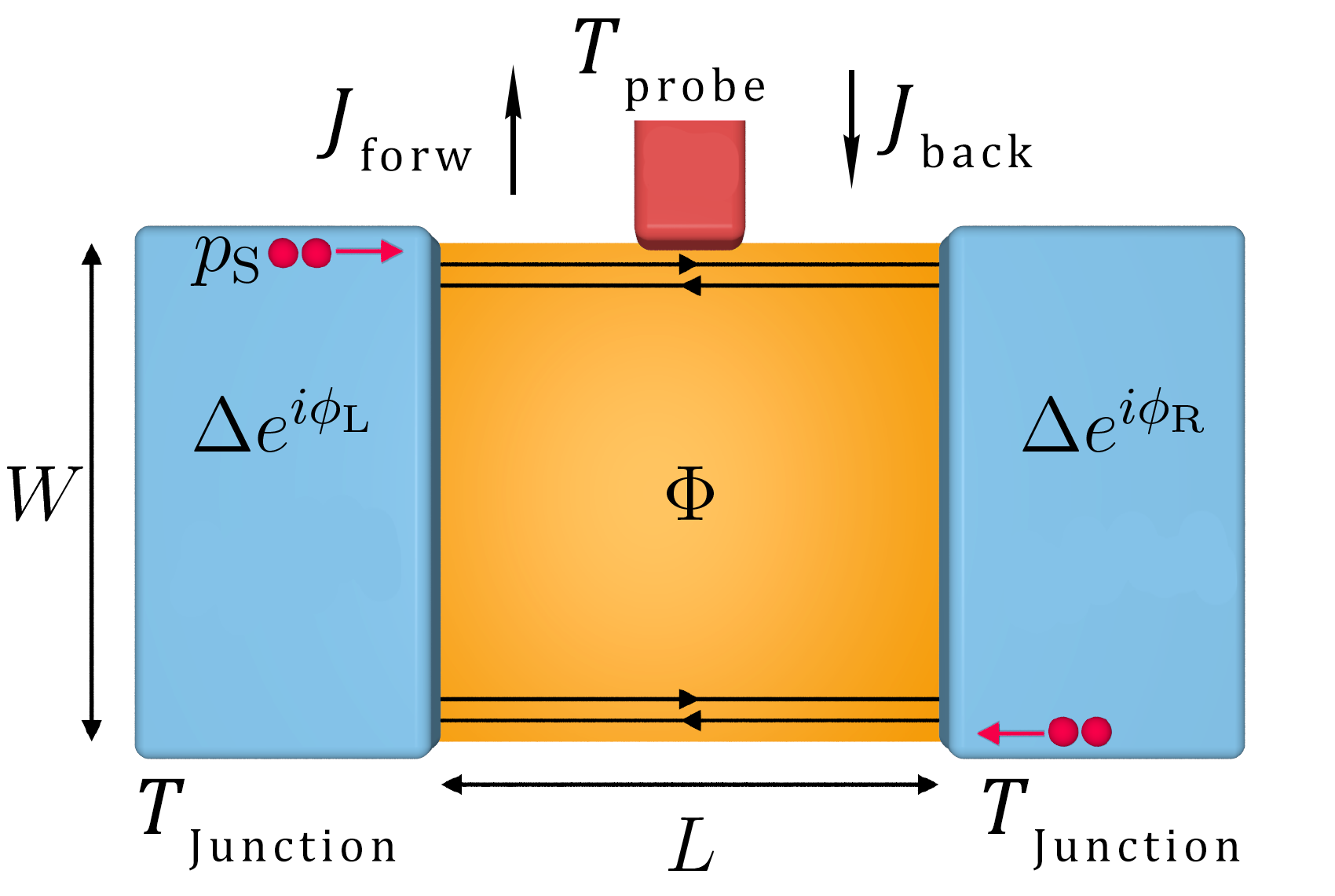}
	\caption{The Topological SQUIPT consists of two superconducting leads (blue) connected by a 2D topological insulator (orange) of width $W$ and length $L$. Transport occurs through the topological edge channels, to which a probe (red) is tunnel coupled. A finite Cooper pair momentum $p_S$ can be induced via a magnetic flux $\Phi$ through the junction. Under forward thermal bias, the junction is hot, while the probe is cold, and vice versa under backward bias.}
	\label{fig:device}
\end{figure}

As a response to recent technological developments, the study of thermal transport in nanostructures has received considerable attention during the past decade. Despite the fact that the flow of heat is not easily controlled, many advances, both theoretical and experimental, have been made, laying the foundation of thermal logic~\cite{Li2012,Bosisio2015,Paolucci2018b,Guarcello2018} and coherent caloritronics~\cite{Meschke2006,Martinez-Perez2014,Fornieri2017,Virtanen2017a,Vischi2018}. Nonlinear thermal elements, such as thermal diodes that conduct heat well in one direction but poorly in the reversed direction are particularly valuable for both thermal logic~\cite{Paolucci2018b} and heat management~\cite{Chen2006}. In this work, we present a thermal diode that is based on the electronic heat flow in a topological insulator (TI) based Josephson junction, cf. Fig.~\ref{fig:device}.

Several proposals and realizations of thermal diodes for electronic heat flow exist, based on quantum dots~\cite{Scheibner2008,Chen2008,Kuo2010,Ruokola2011,Lopez2013}, superconducting elements~\cite{Giazotto2013,Martinez-Perez2013,Fornieri2014,Martinez-Perez2015,Fornieri2015}, the quantum Hall effect~\cite{Granger2009,Nam2013,Sanchez2015}, and TI elements~\cite{Rothe2012,Ren2013,Durst2015,Li2017}. TIs have recently received much interest, as they host interesting physics such as spin-momentum locking, helical edge states, Majorana fermions, which are candidates for quantum computing~\cite{Hasan2010,Qi2011,Ando2013,Tkachov2013} and have peculiar thermal transport properties~\cite{Rizzo2013,Vannucci2015,Ronetti2016,Ronetti2017}. Here, we consider a thermal diode for electronic heat flow which is based on a topological Josephson junction  (TJJ)~\cite{Hart2014a,Hart2017} where superconducting electrodes are connected via the edge states of a two-dimensional TI whose experimental realization can be based on 
2D HgTe/HgCd quantum wells~\cite{Bernevig2006,Konig2007,Konig2008,Roth2009,Buttner2010,Brune2012,Nowack2013}.

Threading a flux through the centre of a 2D TJJ leads to a finite Cooper pair momentum in the superconducting leads, modifying the Andreev reflection amplitudes in the edge channels of the TI~\cite{Tkachov2015}. In the edge channels, left and right moving quasiparticles pick up this momentum shift with opposite sign; an effect which is similar to the Doppler shift for classical waves. The momentum shift modifies the spectrum of left- and right-movers such that one can lift the suppressed of quasiparticle transport through the edge channels~\cite{Bours2018}, sharply increasing the thermal response of the junction. Exploiting this behaviour, a 2D TJJ can be used as an efficient thermal switch~\cite{Martinez-Perez2014,Sothmann2017}. Importantly, the effect requires only small magnetic fields of the order of several mT which leave the helical edge conductance intact~\cite{Pikulin2014}.
 
Previously, we studied the electrical properties of a device consisting of a 2D TJJ with a normal metal probe tunnel coupled to the topological region. In such a Topological version of a Superconducting QUantum Interference Proximity Transistor~\cite{Giazotto2009,Giazotto2011,Meschke2011,Ronzani2014,Dambrosio2015,Virtanen2016,Strambini2016,Vischi2017,Ronzani2017,Ligato2017,Enrico2016,Enrico2017,Jabdaraghi2014,Jabdaraghi2016,Jabdaraghi2017} (or TSQUIPT), see Fig.~\ref{fig:device}, the density of states in the edge channels depends non-trivially on energy, magnetic field and superconducting phase. This allows one to use the TSQUIPT as a sensitive, absolute magnetometer~\cite{Bours2018}. 

Here, we discuss the thermal properties of the TSQUIPT, considering both a probe with a constant density of states (normal metal probe) and a density of states that is linear in energy (graphene probe), and find that it can be used as a thermal diode. The diode's rectification properties derive from the fact that the density of states of the topological junction is implicitly dependent on the temperature, via the induced superconducting gap. The properties of the probe, on the other hand, are temperature independent, which leads to an asymmetric response with respect to temperature.

Due to resonances that appear in the quasi-particle density of states of the junction, the rectification efficiency can be significantly enhanced compared to a conventional NIS junction~\cite{Martinez-Perez2013,Giazotto2013} when using a normal metal probe. As is the case with the electrical properties, the thermal properties of the TSQUIPT depend on the geometry of the junction, the magnetic flux through the junction, and the the superconducting phase difference between the two superconducting leads. As the device is based on a 2D topological insulator, the TSQUIPT could prove useful for heat management, or as part of a cooling scheme in topological insulator based quantum technologies.

This paper is structured as follows. First, we briefly revisit the model used to describe the density of states in the TSQUIPT. We then consider the thermal response by discussing both the linear and non-linear regime and we comment on the thermal rectification performance. We show the dependence on various variables such as the length of the junction, the magnetic flux through the junction, the superconducting phase difference, and temperature, for both a probe with a flat density of states (normal metal) and a linear density of states (graphene). In the final section we summarize the results.

\section{Model}
We consider a Josephson junction formed along the upper edge of a two-dimensional topological insulator, cf. Fig.~\ref{fig:device}. A left (L) and right (R) superconductor are deposited on top of the TI. They induce superconducting correlations in the edge channels via the proximity effect~\cite{Tkachov2013a}. The associated order parameters are characterized by their absolute value $\Delta$ and their phase $\phi_\text{L,R}$.
In between the superconducting electrodes, there is a nonproximitized region of TI with length $L$ forming the actual junction. We assume that the width $W$ of the junction is wide enough such that the overlap between edge states at the upper and lower edge can be neglected. Furthermore, we also neglect any coherent coupling of the edge states via the superconductor which is a valid assumption at high enough temperatures.
The spatial variation of the order parameter between the proximitized and normal region of the edge states is assumed to occur on a length scale much shorter than the superconducting coherence length $\xi_0=\hbar v_{F}/\Delta$, where $v_F$ denotes the Fermi velocity. Furthermore, we neglect proximity effects inside the junction that would require a self-consistent calculation of the order parameter. This is justified as transport occurs only via two edge channels~\cite{Beenakker1991}.
The above approximations finally allow us to write the induced order parameter as $\Delta(x) = \Delta [\Theta(-x-L/2)e^{i\phi_\text{L}} + \Theta(x-L/2)e^{-i\phi_\text{R}}]$, where $\Theta(x)$ is the step function.

The helical edge channels can then be described by the Bogoliubov-de Gennes Hamiltonian
\begin{equation}
	H_\text{BdG}=\left(\begin{array}{cc} h(x) & i\sigma_y \Delta(x) \\ -i\sigma_y \Delta(x)^* & -h^*(x) \end{array}\right).
\end{equation}
The pair potential $\Delta(x)$ couples electronlike and holelike quasiparticles. The single-particle Hamiltonian describing the edge states in the absence of superconductivity and neglecting electron-electron interactions are given by
\begin{equation}
	h(x)=v_{F}\sigma_x\left(-i\hbar\partial_x+\frac{p_{S}}{2}\right)+\sigma_0 \mu,
\end{equation}
with $\sigma_0$ the identity matrix and $\sigma_j$ the Pauli matrices acting on spin space, and the chemical potential $\mu$. We introduced
\begin{equation}
p_S = \frac{\pi \hbar}{L} \frac{\Phi}{\Phi_0} = \frac{\pi \xi_0 \Delta}{v_F L} \frac{\Phi}{\Phi_0}
\label{eq:momentum}
\end{equation}
which denotes the finite momentum of the Cooper-pair condensate along the transport direction and which arises in the presence of a finite magnetic flux $\Phi$ through the junction. While the momentum can be formulated in terms of the magnetic field $B$ and the width $W$ of the junction only, we will always express it and other quantities in terms of the flux $\Phi$ through the junction, and the length $L$ of the junction, assuming the width $W$ is a constant.

\begin{figure}[tb]
\centering
	\includegraphics[width=\columnwidth]{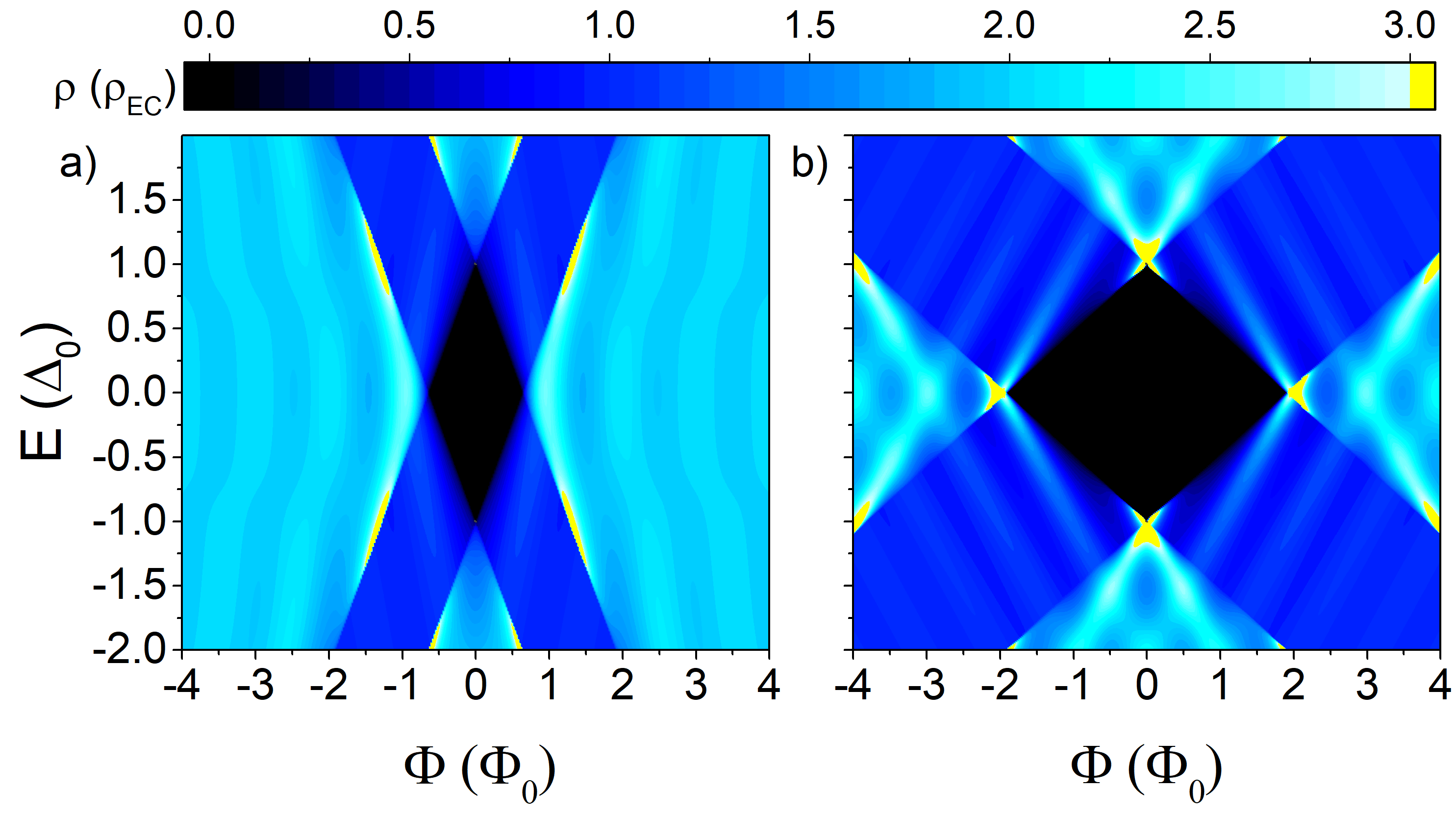}
	\caption{a) The density of states in a junction of $L = \xi_0$, in units of the density of states of the unperturbed density of states of the edge channels $\rho_\text{EC}$, as a function of energy (normalized by $\Delta_0$) and flux (normalized by $\Phi_0$). b) Density of states for a junction of length $L = 3\, \xi_0$. The sub-gap Andreev bound states are omitted, as they do not contribute to stationary thermal transport.}
	\label{fig:dos}
\end{figure}

When Andreev reflection occurs in the presence of a magnetic flux $\Phi$, the left and right moving particles in the edge channels pick up a shift in momentum $p_S/2$ with opposite sign, an effect reminiscent of the Doppler effect. The magnetic field required to reach $\Phi_0$ in a typical junction with $W = 2$ \si{\micro\meter\squared}, and $L=\xi_0 = \hbar v_F / \Delta_0 \approx 600$ nm\cite{Nowack2013, Bocquillon2016a} is $\approx 1$ mT. This field is sufficiently small such that superconductivity is not suppressed and no backscattering is induced or gap is opened in the helical edge channels, thus preserving topological properties of edge channels~\cite{Pikulin2014}. Taking the Doppler effect into account, one can derive the density of states in the upper edge of the junction in units of the edge state density of states $\rho_{EC} = (\pi \hbar v_F)^{-1}$~\cite{Bours2018}
\begin{align}
\label{eq:dos}
\rho_\text{TI} (E) =&\rho_\text{BCS}(E_-)F_-(E_-)\\
+ &\rho_\text{BCS}(E_+)F_+(E_+) \nonumber,
\end{align}
with the normalized superconductor density of states
\begin{equation}
\rho_\text{BCS}(E) = \frac{|E|}{\sqrt{|E|^2 - \Delta^2}}\Theta(|E|-\Delta),
\label{eq:dosBCS}
\end{equation}
and a modulating function
\begin{equation}
\label{eq:interference}
F_\pm(E) = \frac{E^2-\Delta^2}{E^2-\Delta^2 \cos^2(\frac{\phi_u}{2} \pm \frac{E L}{\Delta \xi_0})},
\end{equation}
which originates from quantum interference via Andreev reflections with the TI edge. The energies $E_{\pm} = E \pm\frac{v_{F}p_{S}}{2}$ are the shifted quasiparticle energies~\cite{Bours2018}, see Fig.~\ref{fig:dos}. $\phi_u$ denotes the phase difference along the upper edge, which is defined as $\phi_u = \phi_0 + \pi \Phi / \Phi_0$, where $\phi_0$ is the superconducting phase difference taken in the middle of the junction. The phase difference along the upper edge is considered as we assume a probe is tunnel coupled to the upper side of the 2D TI (see Fig. \ref{fig:device}). Note that the subgap Andreev bound states are omitted in this discussion as they do not contribute to stationary thermal transport. The density of states, $\rho_\text{TI}$ depends on the junction length $L$, the magnetic flux $\Phi$ through the junction, and the superconducting phase difference $\phi_0$ between the two superconducting contacts.

For the sake of readability, we briefly revisit the characteristics of the topological junction before moving to the thermal response of the device. In Fig.~\ref{fig:dos} we present the density of states, which features many details, in a density plot as the function of the flux $\Phi$ and energy $E$ for the superconducting phase difference $\phi_0 = 0$. In the central, black diamond, the quasi--particle spectrum is gapped, and while electrical transport is possible via Andreev bound states (not shown here), thermal transport is blocked completely. For energies above and below the induced gap $\Delta_0$, quasi-particles can be transmitted through both edge channels (top and bottom regions). The same is true at low energies but at high flux, where heat transport because the induced gap is closed by the Doppler shift (left and right regions). In the diagonal stripes, transport is possible through one mode (e.g. only right-moving, or only left-moving particles), while the other remains gapped, due to an interplay between the quasi-particle energy and the Doppler shift, which is opposite for the two counter-propagating channels. We note that, as we do not consider a spin selective probe, it is impossible to exploit these branches for unidirectional heat transport. 
For parameters where transport through the junction is possible, the density of states features modulations that increase in strength for longer junctions. These modulations arise from the merging of Andreev bound states with the continuum and are a consequence of the energy dependence of the electronlie and holelike wave vectors. For a more comprehensive analysis of this effect, see~\cite{Bours2018}.

It is important to note that $\rho_\text{TI}$ is influenced by various parameters. Besides its aforementioned dependence on energy and flux, it depends implicitly on temperature, via the temperature dependence of the superconducting gap $\Delta = \Delta(T)$ as obtained by the BCS theory. As we will see, this dependence will be important for the thermal rectification. Furthermore, as can be seen in Eq.~\eqref{eq:interference} and is shown in Fig.~\ref{fig:dos}, the length of the junction impacts the flux dependence, as it modifies both the slope of the diagonals and the interference pattern. Finally, the superconducting phase difference $\phi_0$ influences the position of the interference pattern, and when not equal to $0$ or $\pi$, breaks the density of states' symmetry with respect to the flux, see again Eq.~\eqref{eq:interference}.

\section{Thermal response}
We now calculate the fully nonlinear thermal tunnel current between the probe and the topological Josephson junction. This is necessary as Onsager symmetry forbids thermal rectification in linear response in the presence of time-reversal symmetry. We use the standard tunneling expression~\cite{Bardeen1961}

\begin{multline}
J_\text{N-TI}= \\ \frac{1}{e^2 R_{T}} \int_{-\infty}^{+\infty} E \rho_\text{TI}(E) \rho_\text{P} (E) [f_\text{P}(E)-f_\text{TI}(E)] dE.
\label{eq:J}
\end{multline}
For the density of states of the probe $\rho_\text{P} (E)$, we consider two different scenarios: a normal metal probe and a graphene probe. For the normal metal probe, the density of states is assumed to be energy independent, $\rho_\text{P} = \rho_0$, in which case $R_T$ is the tunnel resistance between the probe and the junction. For the graphene probe the density of states is energy dependent: $\rho_\text{P} = 2 |E| / (\pi \hbar^2 v_F^2 )$, and $R_T$ represents a resistance factor associated with the transparency of the tunnel barrier between the probe and the junction. The density of states in the junction, $\rho_\text{TI}(E)$, is given by Eq.~\ref{eq:dos}. The Fermi functions $f_\text{P}(E)$ and $f_\text{TI}(E)$ are associated with the probe and the junction respectively, with $f_\text{P,TI}(E) = (1+e^{E/k_B T_{P,TI}})^{-1}$. We assume that the chemical potential is kept constant everywhere: $\mu_\text{S} = \mu_\text{P} = 0$, with $\mu_\text{S}$ and $\mu_\text{P}$ the chemical potential in the superconducting leads and the probe respectively, to avoid an extra dissipative charge current flowing in the junction. To ascertain that the topological insulator remains in thermal equilibrium with the superconducting leads, and the tunnelling approximation is valid, it is necessary that $R_T \gg R_K$, with $R_K$ the von Klitzing resistance. Only when the thermal conductance between the probe and the junction is much smaller than the thermal conductance between the topological edge channels and the superconducting leads, one can safely assume thermal equilibrium in the topological junction.

\begin{figure}[btp]
\centering
	\includegraphics[width=\columnwidth]{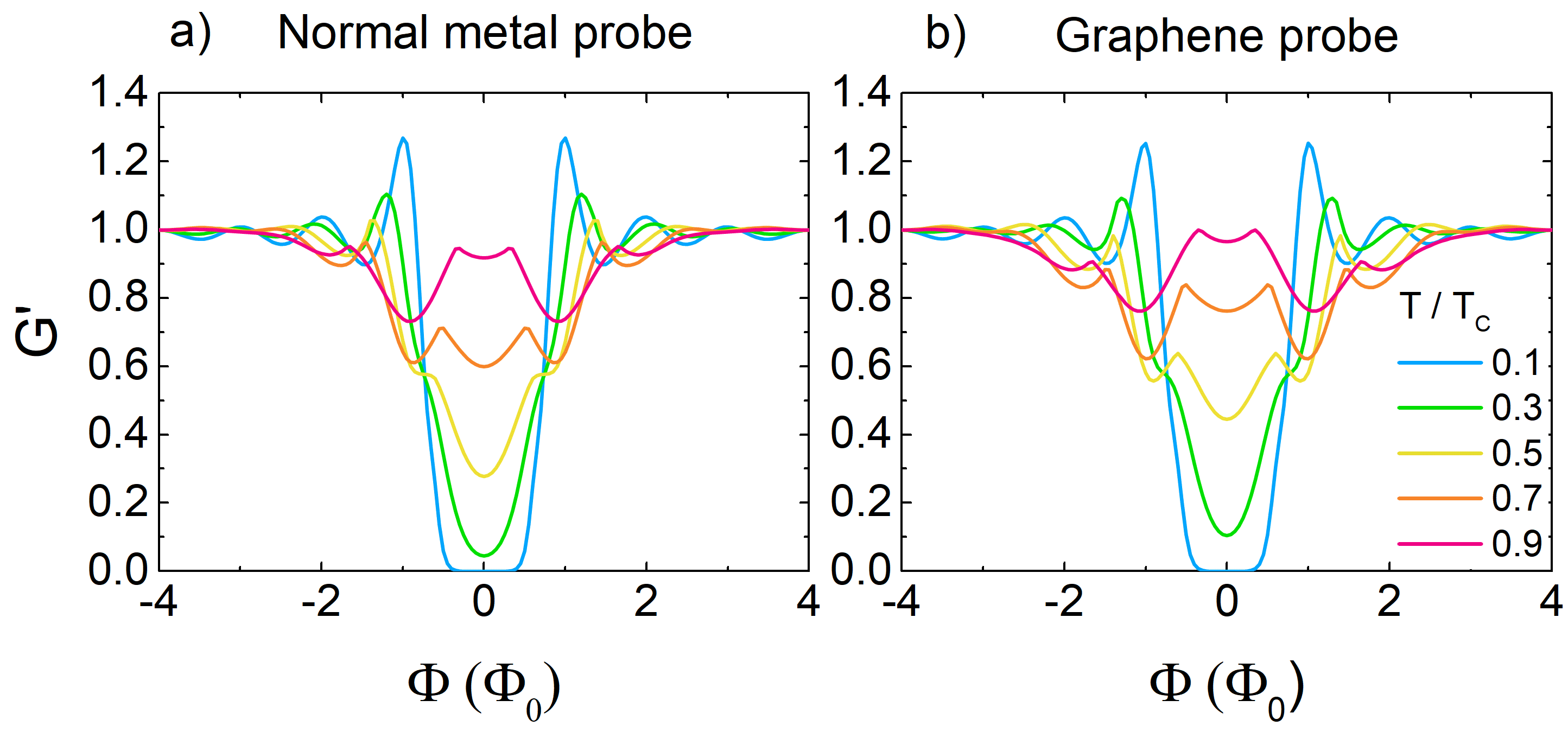}
	\caption{The normalized thermal conductance in the linear regime $G'(\Phi)$ as a function of the magnetic flux through the junction at various temperatures. a) Conductance for a probe with a flat density of states, i.e. a normal metal and b) conductance for a probe with a linear density of states, i.e. a graphene probe. In both cases the junction length is $L = \xi_0$ and superconducting phase difference $\phi_0 = 0$.}
	\label{fig:G}
\end{figure}

\subsection{Linear regime}
Assuming the device is operated in the linear regime, i.e. the thermal gradient $\delta T = |T_\text{TI}-T_\text{P}| \ll T$, we can define the thermal conductance as:
\begin{equation}
J_T=G_T(\Phi) \delta T,
\end{equation}
where
\begin{equation}
G_T(\Phi)= \frac{1}{4 e^2 R_{T}} \int_{-\infty}^{+\infty} \frac{E^2}{T^2}\frac{ \rho_\text{TI}(E) \rho_\text{P}(E)}{\cosh^2\frac{E}{2\kBT}}~dE.
\end{equation}

\begin{figure}[btp]
\centering
	\includegraphics[width=\columnwidth]{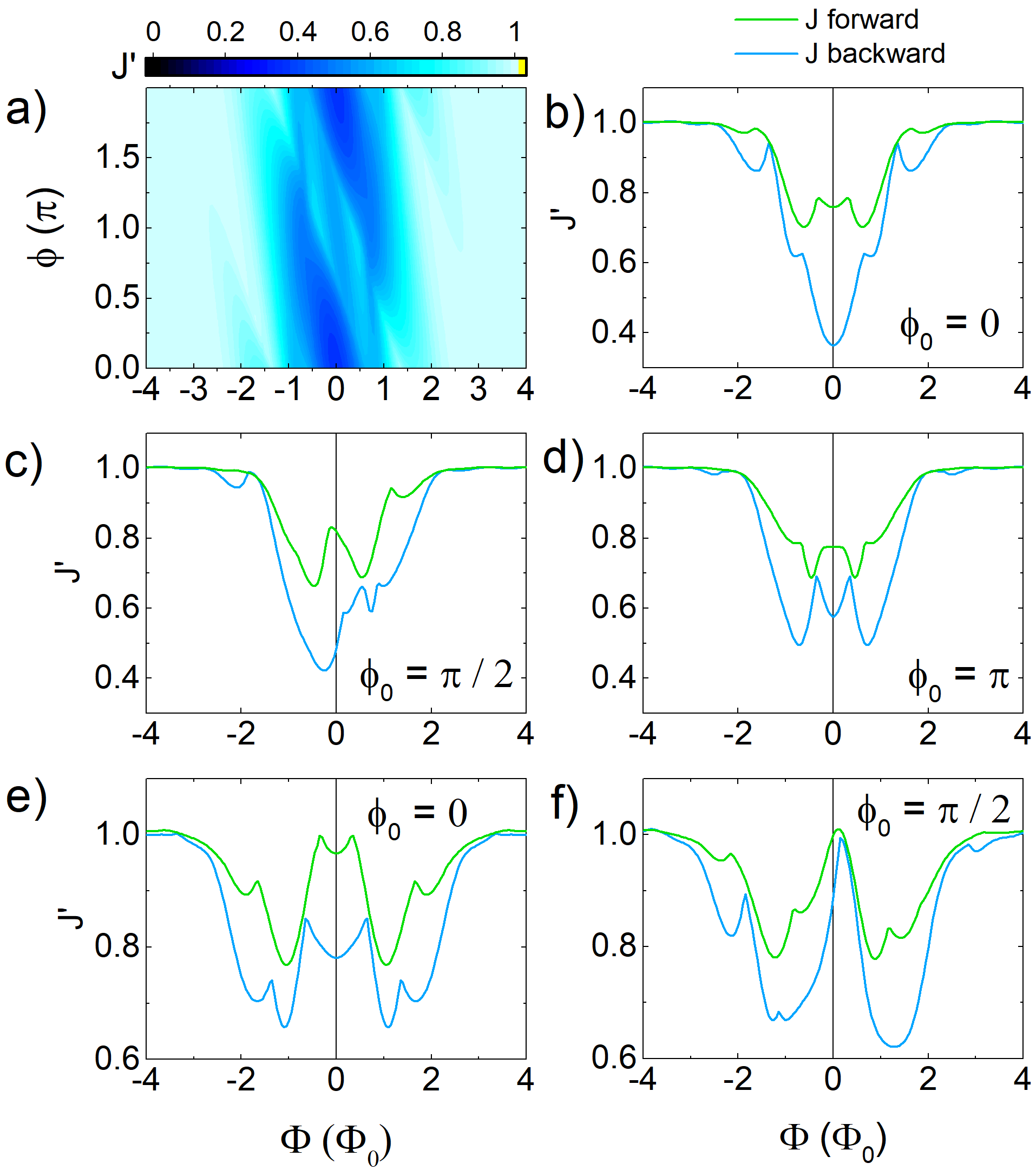}
	\caption{a) The normalized heat current $J'(\Phi)$, in the forward configuration, with a hot junction ($T = 0.9\, T_\text{C}$) and a cold probe ($T = 0.1\, T_\text{C}$). Junction length is $L = \xi_0$, and the heat current is plotted as a function of the flux through the junction (normalized by $\Phi_0$) and the superconducting phase difference $\phi_0$ in the middle of the junction. b) The normalized forward and backward heat current as a function of the normalized flux for a junction of $L = \xi_0$, with $T_\text{Hot}$ = 0.9 $T_\text{C}$ and $T_\text{Cold}$ = 0.1 $T_\text{C}$, and superconducting phase difference $\phi_0 = 0$ b) $\phi_0 = \pi /2$ and c) $\phi_0 = \pi$. e) and f) show the normalized heat current for a graphene probe, at $\phi_0 = 0$ and $\phi_0 = \pi/2$ respectively, for the same parameters as previously.}
	\label{fig:J}
\end{figure}

\begin{figure*}[bt]
\centering
	\includegraphics[width=\textwidth]{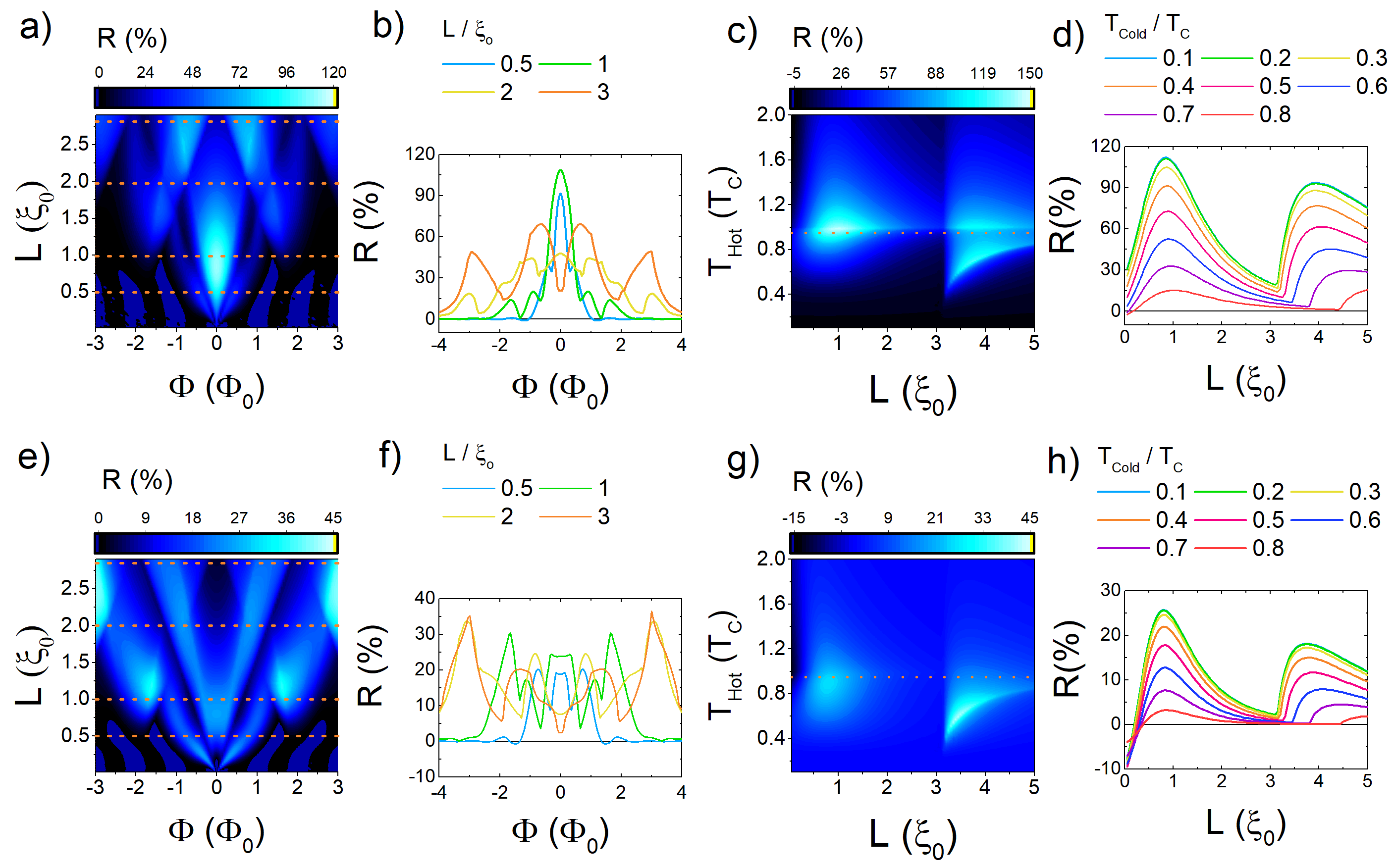}
	\caption{a) - d) correspond with a normal metal probe. a) A density plot of the rectification coefficient as defined in eq.~\ref{eq:rectification}, as a function of the normalized flux and the junction length b) $\mathcal{R}$ versus flux for junctions of four different lengths. The position of the cuts is indicated by the orange dotted lines and arrows in a). c) A density plot of the rectification coefficient as a function of the length of the junction and the $T_\text{Hot}$. $T_\text{Cold} = 0.1\, T_\text{C}$, $\Phi = \phi_0 = 0$ d) $\mathcal{R}$ versus flux for various values of $T_\text{Cold}$, $T_\text{Hot} = 0.9\, T_\text{C}$, and $\Phi = \phi_0 = 0$. The position of the light blue line ($T=0.1\, T_\text{C}$) is indicated by the dotted orange line in c). e) - h) Like a) - d), but with a graphene probe.}
	\label{fig:RvsEverything}
\end{figure*}

\begin{figure}[hbtp]
\centering
	\includegraphics[width=\columnwidth]{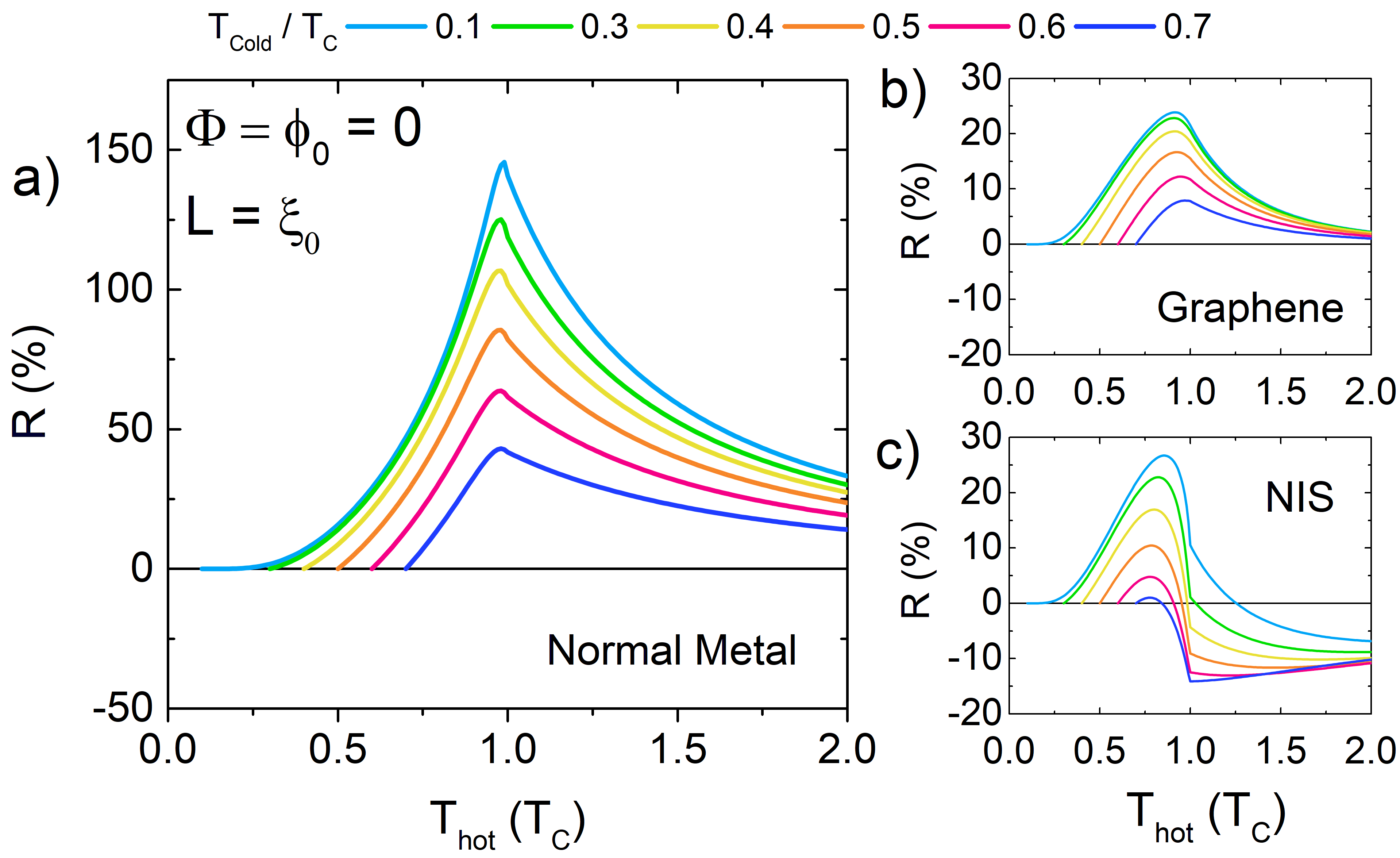}
	\caption{a) The rectification coefficient $\mathcal{R}$, as defined in eq. ~\ref{eq:rectification}, for a normal metal probe, as a function of $T_\text{Hot}$, for various values of $T_\text{Cold}$. Junction $L=\xi_0$, and $\Phi = \phi_0 =0$. b) The rectification coefficient for a graphene probe. c) The thermal rectification of a NIS junction is shown for comparison.}
	\label{fig:RvsT}
\end{figure}

The thermal conductance, like the thermal current, directly reflects the matching between the density of states of the probe and the topological junction. It depends on the temperature $T$, length of the junction $L$, the flux through the junction $\Phi$, and the superconducting phase bias $\phi_0$, as follows from Eqs.~(\ref{eq:dos},~\ref{eq:dosBCS},~\ref{eq:interference}).

In Fig.~\ref{fig:G} we show the thermal conductance of the TSQUIPT as a function of the flux through the junction normalized by the conductance at high flux, $G'(\Phi) = G (\Phi) / G_\infty$ where $G_\infty = G(\Phi \rightarrow \infty)$. For low temperatures, the heat flow through the device is completely suppressed by the presence of the superconducting gap. However, by applying a flux one can close the quasi-particle gap, allowing heat transport to occur. For temperatures closer to $T_\text{C}$, the size of the gap, and thus the suppression of the conductance decreases, however a modulation is clearly visible up to $0.9\,T_\text{C}$. For magnetic fluxes larger than a few flux quanta, the conductance reaches a constant value and the modulation decays to zero. While the thermal conductance for both probes looks very similar, it is worthwhile to note that the thermal conductance increases linearly with temperature for a normal probe, but super-linear for a graphene probe (not shown here).

\subsection{Non-linear regime}
In Fig.~\ref{fig:J} a) we show the normalized thermal current $J'(\Phi) = J(\Phi) / J_\infty$ with $J_\infty=J(\Phi\to\infty)$ in the forward configuration (hot junction, cold probe) as a function of the flux $\Phi$ and superconducting phase bias $\phi_0$, for a normal metal probe. When $\phi_0\neq 0,\pi$, an asymmetry in the flux dependence of the heat current is induced. In Fig.~\ref{fig:J} b) - d), we present the normalized forward (hot junction, cold probe, green line) and backward (hot probe, cold junction, blue line) heat currents, for various values of the superconducting phase difference $\phi_0$ and a normal metal probe, as a function of the flux. When a large magnetic flux is applied, heat transport is fully enabled by the Doppler effect. As the impact of the gap diminishes, so does the difference between the forward and backward heat currents. Fig.~\ref{fig:J} e) and f) show the forward and backward heat currents for the case of a graphene probe. The qualitative behavior of the graphene probe device is similar to the normal probe device, as the dependence on flux and phase mainly originates from the junction, and not from the nature of the probe.

The difference between the forward and backward heat current, and thus the diode's rectification properties, stem from the fact that the density of states in the probe is temperature independent, while that of the junction is modulated by temperature via the dependence of the induced superconducting gap. This results in an asymmetry in forward and backward heat flow, which can be understood from Eq.~\ref{eq:J}. In the absence of the temperature dependence in $\rho_\text{TI}(E)$, the interchange of temperatures affects only the Fermi functions, leading solely to a change of sign of the thermal current. The temperature dependence of the gap is thus fundamental for the asymmetric thermal response of the thermal current which will be discussed in detail later.

If the junction heats up to temperatures approaching the critical temperature $T_\text{C}$, the magnitude of the induced gap decreases rapidly. This allows for an increased heat flow from the junction to the probe. Conversely, if the probe is hot, but the junction is cold, the presence of the gap prevents heat flow for temperatures up to $T \approx T_\text{C}$. As the density of states of the graphene probe is linear in energy, it gives a greater weight to states that are higher in energy when calculating the tunnel current, and the discrepancy between the forward and backward heat current is reduced.

\subsection{Thermal rectification}
Finally, to quantify the effectiveness of the thermal rectification, we define the relative rectification coefficient
\begin{equation}
\mathcal{R (\%)} = 100 \cdot \frac{|J_+| -|J_-|}{|J-|},
\label{eq:rectification}
\end{equation}
where $J_+$ denotes the forward heat current, corresponding to a hot junction and a cold probe, and $J_-$ is the backward current, where the temperatures are reversed.

As the resonances in the junction influence the density of states above the gap and by extension the heat transport, the length of the junction and the flux through the junction have a strong effect on the thermal properties. In Fig.~\ref{fig:RvsEverything} a) we show the dependence of the rectification coefficient on the length of the junction and the flux through the junction for the normal probe. Panel b) shows the rectification coefficient for four different lengths, corresponding to cuts of panel a). For the shorter junctions, the rectification has a sharp peak around zero flux, while for the longer junctions, the rectification at zero flux becomes a local minimum, although it partly recovers at finite values of the flux. Fig.~\ref{fig:RvsEverything} c) is a density plot of the rectification as a function of the junction length and $T_\text{Hot}$, while $T_\text{Cold} = 0.1 \, T_\text{C}$ is kept constant. There are two notable features: firstly, it is clear that for maximum rectification, $T_\text{Hot}$ should be close to $T_\text{C}$, independent of the length. Secondly, as can also be seen in panel d), the rectification peaks at $L = \xi_0$, and then diminishes until $L \approx 3\,\xi_0$, after which it largely recovers before decaying again. This oscillation of the rectification with respect to the junction length is a consequence of the position of the maxima in the density of states of the junction, that arise from quantum-mechanical interference. 

Fig.~\ref{fig:RvsEverything} e) - h) show the behaviour of the rectification with a graphene probe. Contrary to before, the maximum of $\mathcal{R}$ now moves outwards to higher values of the $\Phi$ as the length of the junction increases. However, the value of the maximum stays between $\approx 25 \%$ and $\approx 40$ \%, depending on the various parameters. Furthermore, for a graphene probe connected to very short junctions, the rectification coefficient will be negative, especially when $T_\text{Cold}$ is low, which indicates a reversal of the device's expected response. This behavior is similar to that of a NIS junction (cf. Fig.~\ref{fig:RvsT} c)), however the effect is enhanced as the graphene density of states gives a greater weight to high energy states.

The maximum rectification coefficient is obtained using a normal metal probe. Following the discussion of Fig.~\ref{fig:RvsEverything}, we find that maximum value of the rectification coefficient $\mathcal{R} \approx 145 \%$ is found for a junction of length $\xi_0$, with $T_\text{Cold} \rightarrow 0$ and $T_\text{Hot} \rightarrow T_\text{C}$, $\Phi = 0$ and $\phi_0 = 0$. At a length of $3\, \xi_0$ and $\Phi = 0$, the efficiency is minimum, with a value of $\mathcal{R}\approx 45 \%$, although it recovers to $\mathcal{R} \approx 70 \%$ at $|\Phi| \approx 0.7\, \Phi_0$. When considering a graphene probe, the optimum quantity of flux increases with the device length, although the maximum value never exceeds $\mathcal{R} \approx 40\%$.

Fig.~\ref{fig:RvsT} shows the rectification coefficient as a function of T$_\text{Hot}$, for various values of $T_\text{Cold}$, for a junction of $L=\xi_0$. At this length, the rectification is maximum for $\Phi = \phi = 0$, but this is not necessarily true for junctions of a different length. The rectification peaks just below $T_\text{C}$, and decreases in a constant fashion as $T_\text{Cold}$ increases, which demonstrates that the maximum rectification factor is obtained with $T_\text{Cold}$ at the lowest possible temperature. When using a graphene probe (Fig.~\ref{fig:RvsT} b)), the rectification coefficient is reduced, with a maximum value and shape that is reminiscent of the NIS diode. However, the temperature range of positive rectification is increased with respect to that of the NIS junction, especially for temperatures above the critical temperature. The graphene probe, which reduces the effect of the junction density of states, can thus be seen as an intermediate between the NIS junction and the TSQUIPT with normal metal probe in terms of thermal rectification performance.

\begin{figure}[tp]
\centering
	\includegraphics[width=\columnwidth]{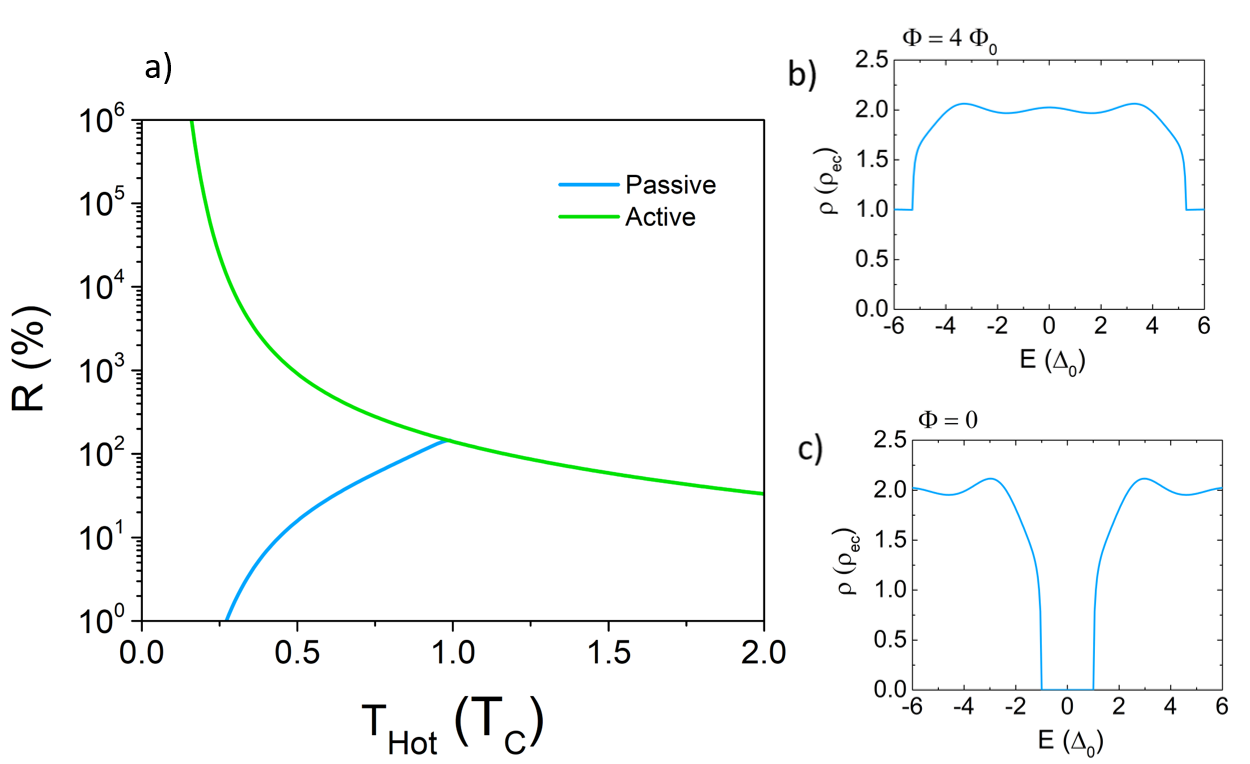}
	\caption{a) The rectification coefficient of the diode for $T_\text{Cold} = 0.1\, T_\text{C}$, a junction of length $L = \xi_0$, and $\Phi = \phi_0 = 0$ (passive mode, identical to the blue line in fig.~\ref{fig:RvsT}) versus the rectification coefficient when the diode is open in the forward configuration at $\Phi = 4 \, \Phi_0$, but closed in backward configuration with $\Phi = 0$ (active mode). b) The density of states of the junction at $T = 0.1 \, T_\text{C}$, $L = \xi_0$ and $\Phi = 4 \Phi_0$. In this case the quasiparticles can flow freely from the junction into the normal metal probe. c) The same as b) except now $\Phi = 0$. The induced gap suppresses the heat flow from the probe to the junction.}
	\label{fig:RActive}
\end{figure}

To further increase the rectification coefficient, one could consider replacing the normal metal tunnel coupled probe with a superconducting probe, in this case, an enhancement similar to the one present in~\cite{Martinez-Perez2013} can be expected, although the extra Josephson coupling should also be taken into account.

Alternatively, one could think to exploit the switching effect of the junction where the quasi--particle gap can be closed by applying a magnetic flux~\cite{Sothmann2017}. Closing the quasi-particle gap enhances the thermal response of the junction. Hence, when closing the gap by applying a large flux in the forward configuration (hot junction, cold probe), the quasi-particle current and thus the heat flow through the device is increased, see the density of states presented in Fig.~\ref{fig:RActive} b). On the other hand, keeping the gap open i.e. not applying a magnetic flux, in the backward configuration (cold junction, hot probe) ensures a minimal back flow, see the density of states shown in Fig.~\ref{fig:RActive} c). Thus, by using the device dependence on the magnetic flux, one can implement an active rectification scheme. Since the heat flow is exponentially suppressed with the channels closed and $k_B T \ll \Delta_0$, rectification is strongly enhanced below $T_\text{C}$, as shown in Fig.~\ref{fig:RActive} a).

\section{Discussion and conclusion}
We have shown that the topological SQUIPT can be utilized as a thermal diode, and how the rectification efficiency depends on various device parameters such as the length $L$, the magnetic flux through the topological Josephson junction $\Phi$ and the phase difference between the superconducting leads $\phi_0$, and for probes with a flat and a density of states that is linear in energy. A passive rectification coefficient of up to $\approx 145 \%$ is reached under optimal conditions, when using a normal metal probe, which is high with respect to comparable designs such as NIS or SIS' junctions. An interesting property of the TSQUIPT is that it offers control over the superconducting gap via a small external magnetic field, which can be used to open the junction to heat flow when needed. Exploiting this effect in an active rectification scheme, the rectification can be strongly enhanced for temperatures below $T_\text{C}$. The proposed device is a promising tool for managing heat flows in temperature sensitive 2D topological insulator based quantum technologies.

\begin{acknowledgments}
E. S., A. B. and F. G. acknowledge financial support from ERC grant agreement no. 615187 – COMANCHE.  A.B. and F.G. acknowledge the Royal Society though the International Exchanges between the UK and Italy (grant
IESR3 170054).

The work of F. G. was partially funded by the Tuscany Region under the FARFAS 2014 project SCIADRO. M. C. and A.B. acknowledge support from the CNR-CONICET cooperation programme ``Energy conversion in quantum, nanoscale, hybrid devices''. E.M.H. and L.W.M.   acknowledge   financial   support   from the German Science Foundation (Leibniz Program,  SFB1170  ``ToCoTronics'')  and  the  Elitenetzwerk  Bayern  program  ``Topologische Isolatoren''. M. C. acknoledges support from the Quant-Era project ``SuperTop''. 

We acknowledge financial support from the Ministry of Innovation NRW via the ``Programm zur Förderung der Rückkehr des hochqualifizierten Forschungsnachwuchses aus dem Ausland''.
\end{acknowledgments}


%

\end{document}